\documentclass[twocolumn,trackchanges]{aastex62}
\graphicspath{{./}{figures/}}

\received{Soon}
\revised{After that}
\accepted{\today}
\submitjournal{ApJ}

\shorttitle{SN~2017cbv Nebular Spectroscopy}
\shortauthors{Sand et al.}

\begin{document}

\title{Nebular Spectroscopy of the `Blue Bump' Type Ia Supernova 2017cbv}

\correspondingauthor{David J. Sand}
\email{dsand@as.arizona.edu}

\author{D. J. Sand}
\affil{Department of Astronomy/Steward Observatory, 933 North Cherry Avenue, Rm. N204, Tucson, AZ 85721-0065, USA}

\author{M. L. Graham}
\affiliation{Department of Astronomy, University of Washington, Box 351580, Seattle, WA 98195, USA}

\author{J. Boty\'{a}nszki}
\affiliation{Physics Department, University of California, Berkeley, CA 94720, USA}

\author{D. Hiramatsu}
\affiliation{Department of Physics, University of California, Santa Barbara, CA 93106-9530, USA}
\affiliation{Las Cumbres Observatory, 6740 Cortona Dr, Suite 102, Goleta, CA 93117-5575, USA}

\author{C. McCully}
\affiliation{Department of Physics, University of California, Santa Barbara, CA 93106-9530, USA}
\affiliation{Las Cumbres Observatory, 6740 Cortona Dr, Suite 102, Goleta, CA 93117-5575, USA}

\author{S. Valenti}
\affiliation{Department of Physics, University of California, 1 Shields Avenue, Davis, CA 95616-5270, USA}

\author{G. Hosseinzadeh}
\affiliation{Department of Physics, University of California, Santa Barbara, CA 93106-9530, USA}
\affiliation{Las Cumbres Observatory, 6740 Cortona Dr, Suite 102, Goleta, CA 93117-5575, USA}

\author{D. A. Howell}
\affiliation{Department of Physics, University of California, Santa Barbara, CA 93106-9530, USA}
\affiliation{Las Cumbres Observatory, 6740 Cortona Dr, Suite 102, Goleta, CA 93117-5575, USA}


\author{J. Burke}
\affiliation{Department of Physics, University of California, Santa Barbara, CA 93106-9530, USA}
\affiliation{Las Cumbres Observatory, 6740 Cortona Dr, Suite 102, Goleta, CA 93117-5575, USA}

\author{R. Cartier}
\affiliation{Cerro Tololo Inter-American Observatory, National Optical Astronomy Observatory, Casilla 603, La Serena, Chile}

\author{T. Diamond}
\affiliation{Goddard Space Flight Center, 8800 Greenbelt Rd, Greenbelt, MD 20771, USA}

\author{E.~Y.  Hsiao}
\affiliation{Department of Physics, Florida State University, Tallahassee, FL 32306, USA}

\author{S.W. Jha}
\affiliation{Department of Physics and Astronomy, Rutgers, the State University of New Jersey, 136 Frelinghuysen Road, Piscataway, NJ 08854, USA}

\author{D. Kasen}
\affiliation{Physics Department, University of California, Berkeley, CA 94720, USA}
\affiliation{Astronomy Department and Theoretical Astrophysics Center, University of California, Berkeley, CA 94720, USA}
\affiliation{Nuclear Science Division, Lawrence Berkeley National Laboratory, Berkeley, CA 94720, USA}

\author{S. Kumar}
\affiliation{Department of Physics, Florida State University, Tallahassee, FL 32306, USA}

\author{G.~H. Marion}
\affiliation{Department of Astronomy, The University of Texas at Austin, 1 University Station C1400, Austin, TX 78712-0259, USA}


\author{N. Suntzeff}
\affiliation{George P. and Cynthia Woods Mitchell Institute for Fundamental Physics and Astronomy, Department of Physics and Astronomy, Texas A\&M University, College Station, TX, 77843, USA}

\author{L. Tartaglia}
\affiliation{Department of Astronomy and The Oskar Klein Centre, AlbaNova University Center, Stockholm University, SE-106 91 Stockholm, Sweden}

\author{J. C. Wheeler}
\affiliation{Department of Astronomy, The University of Texas at Austin, 1 University Station C1400, Austin, TX 78712-0259, USA}

\author{S. Wyatt}
\affil{Department of Astronomy/Steward Observatory, 933 North Cherry Avenue, Rm. N204, Tucson, AZ 85721-0065, USA}


\begin{abstract}

We present nebular phase optical and near-infrared spectroscopy of the Type Ia supernova (SN) 2017cbv.  The early light curves of SN~2017cbv showed a prominent blue bump in the $U$, $B$ and $g$ bands lasting for $\sim$5 d.  One interpretation of the early light curve was that the excess blue light was due to shocking of the SN ejecta against a nondegenerate companion star -- a signature of the single degenerate scenario. If this is the correct interpretation, the interaction between the SN ejecta and the companion star could result in significant H$\alpha$ (or helium) emission at late times, possibly along with other species, depending on the companion star and its orbital separation.  A search for H$\alpha$ emission in our +302 d spectrum yields a nondetection, with a $L_{H\alpha}$$<$8.0$\times$10$^{35}$ erg/s  (given an assumed distance of $D$=12.3 Mpc), which we have verified by implanting simulated H$\alpha$ emission into our data. We make a quantitative comparison to models of swept-up material stripped from a nondegenerate companion star, and limit the mass of hydrogen that might remain undetected to $M_{\rm H} < 1 \times 10^{-4}$ $\rm M_{\odot}$. A similar analysis of helium star related lines yields a $M_{\rm He} < 5 \times 10^{-4}$ $\rm M_{\odot}$.  Taken at face value, these results argue against a nondegenerate H or He-rich companion in Roche lobe overflow as the progenitor of SN 2017cbv. Alternatively, there could be weaknesses in the envelope-stripping and radiative transfer models necessary to interpret the strong H and He flux limits.

\end{abstract}


\section{Introduction} \label{sec:intro}

Type Ia supernovae (SNe~Ia) result from the thermonuclear explosion of carbon-oxygen (C-O) white dwarfs \citep{Hoyle60}, but despite their critical use as standardizable candles to measure the expansion history of the Universe, their explosion mechanisms and progenitor systems are still unclear \citep[e.g.,][for a recent review]{Maoz14}.

SN~Ia progenitor systems fall under two broad categories -- the single degenerate (SD) and double degenerate (DD) scenarios.  In the SD scenario, the C-O white dwarf has a nondegenerate companion star \citep{Whelan73}, while the DD scenario has a second degenerate companion in the system \citep{Iben84,Webbink84}.  Within these two broad categories, several explosion mechanisms (i.e. how the thermonuclear explosion is triggered) are being actively studied.

The presence of a nondegenerate companion star can be revealed through  observational signatures which we touch on here, although not exhaustively.  Theoretical models predict a ``blue bump'' in the UV-optical light curve in the days after explosion due to the SN ejecta impacting the nondegenerate companion \citep[e.g.,][]{Kasen10}. This signature has been reported in a handful of instances \citep{Cao15,Marion16,Hosseinzadeh17}, although not all works have interpreted  these light curve features as companion-shock interaction \citep[e.g.,][and see also the early, redder excess in \citealt{Jiang17}]{Miller18}. Early light curve excesses might also result from extended $^{56}$Ni distributions \citep[possibly in combination with circumstellar material interaction;][]{Piro16}, which may result in unusual early optical spectra \citep[][]{Jiang17}. 

Another key prediction of the SD scenario is that material stripped from the companion star is swept up by the SN Ia ejecta, detectable as a relatively narrow emission line of hydrogen or helium at late times \citep[e.g.,][among others]{Marietta00,Mattila05,Pan10,Pan12,Liu12,Liu13,Lundqvist13,Boty18}. The models for the emission from stripped material have considerable diversity in the strength and shape of the emission line, and depend on the details of the companion type and binary separation. Here we will rely on the latest modeling by \cite{Boty18}, but provide interpretations based on previous models as well. 
Despite their promise, definitive late-time narrow emission lines have not been seen in standard SN Ia   \citep{Mattila05,Leonard07,Shappee13,Lundqvist15,Maguire16,Graham17,Shappee18}. 


The nearby Type Ia SN~2017cbv, which exhibited the clearest early light curve ``blue bump" to date and is a strong candidate to originate from a SD progenitor \citep[see][for details]{Hosseinzadeh17}, presents us with an opportunity to search for nebular emission lines associated with companion star stripping in order to cross-check the early light curve results. 
SN~2017cbv was discovered on 2017 March 10 (UT) with a magnitude of $R$$\sim$16 mag by the Distance Less Than 40 Mpc survey \citep[DLT40;][]{Tartaglia18}.  Within hours of discovery, follow-up imaging \citep{DLT17u_disc} and spectroscopy \citep{DLT17u_firstspec} were obtained, confirming the source to be a young Type Ia SN.  After reaching an apparent $B$-band maximum of 11.72 mag on MJD=57841.07, SN~2017cbv exhibited a normal light curve decline rate of $\Delta$m$_{15}$($B$)=1.06 mag \citep{Hosseinzadeh17}.  The SN occurred in the outskirts of the nearby galaxy NGC~5643, which has a reported distance modulus of $\mu$=31.14$\pm$0.40 mag, or $D$=$16.9 \pm 3.1$~Mpc \citep{1988ngc..book.....T}.  This distance is quite uncertain, and we revisit it using the SN~2017cbv light curve itself in the next section.
High resolution spectroscopy did not detect time variable narrow line emission associated with SN~2017cbv \citep{Ferretti17}, which limits the radii and amount of any circumstellar material in the vicinity of the progenitor system, but does not formally rule out the model presented by \citet{Hosseinzadeh17}.

\begin{table*}
\begin{center}
\begin{tabular}{cccccccc}
\hline
\hline
Observation & Phase  & Instrument & Grating/Grism and Central & Slit            & Filter & Exposure & Airmass  \\ 
Date (UT)   & [days] &            & Wavelength [nm]           & Width [\arcsec] &        & Time [s] &  \\
\hline
2018 Jan 09 & +286.3 & GMOS        & B600/450.0     & 1.0\arcsec & --    & 2$\times$450  & 1.33 \\ 
            &        &             & R400/750.0     & 1.0\arcsec & OG515   & 1$\times$450  & 1.33 \\ 
2018 Jan 10 & +287.2 & Flamingos-2 & JH\_G5801/1400 & 0.72\arcsec & JH       & 6$\times$240 & 1.38\\
2018 Jan 25 & +302.3 & GMOS        & R400/750.0     & 1.5\arcsec & OG515   & 4$\times$1200 & 1.23 \\
\hline
\end{tabular} 
\caption{Nebular spectroscopy log of SN~2017cbv.}
\label{tab:gmos}
\end{center}
\end{table*}

Here we present nebular-phase optical and near-infrared spectroscopy of SN~2017cbv, and search for the emission lines of hydrogen (or helium) rich material from a putative non-degenerate companion. In Section \ref{sec:obs} we present our photometric and spectroscopic observations of SN~2017cbv, along with a distance estimate based on the light curve. In Section \ref{sec:ana} we search for narrow emission lines from material stripped off the companion star and estimate the flux limits of our non-detections based on the signal-to-noise ratio of our data, and an implanted emission line feature. In Section \ref{sec:masslimits} we use published models for stripped material to constrain the mass of hydrogen and helium in the progenitor system of SN~2017cbv, and discuss the physical implications of our results. We summarize and conclude in Section~\ref{sec:conclude}. A flat cosmology with $\Omega_M=0.3$,  $\Omega_{\Lambda}=0.7$ and $H_0$=72 km s$^{-1}$ Mpc$^{-1}$ is assumed.

\section{Observations} \label{sec:obs}
We present new photometry and late-time spectroscopy of SN~2017cbv.

\subsection{Photometry} \label{ssec:obs_phot}

\begin{figure}
\includegraphics[width=8cm]{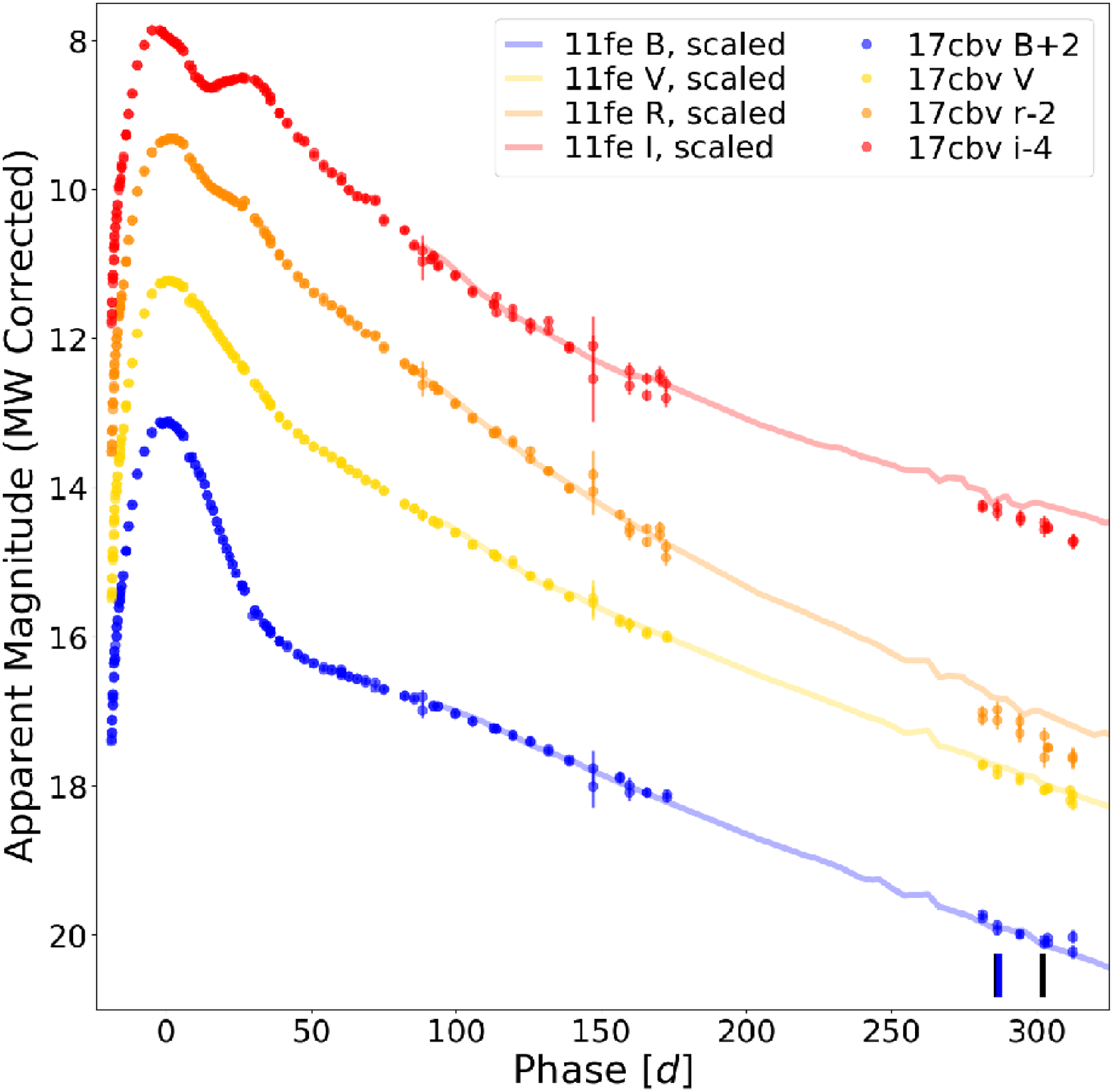}
\includegraphics[width=8.2cm]{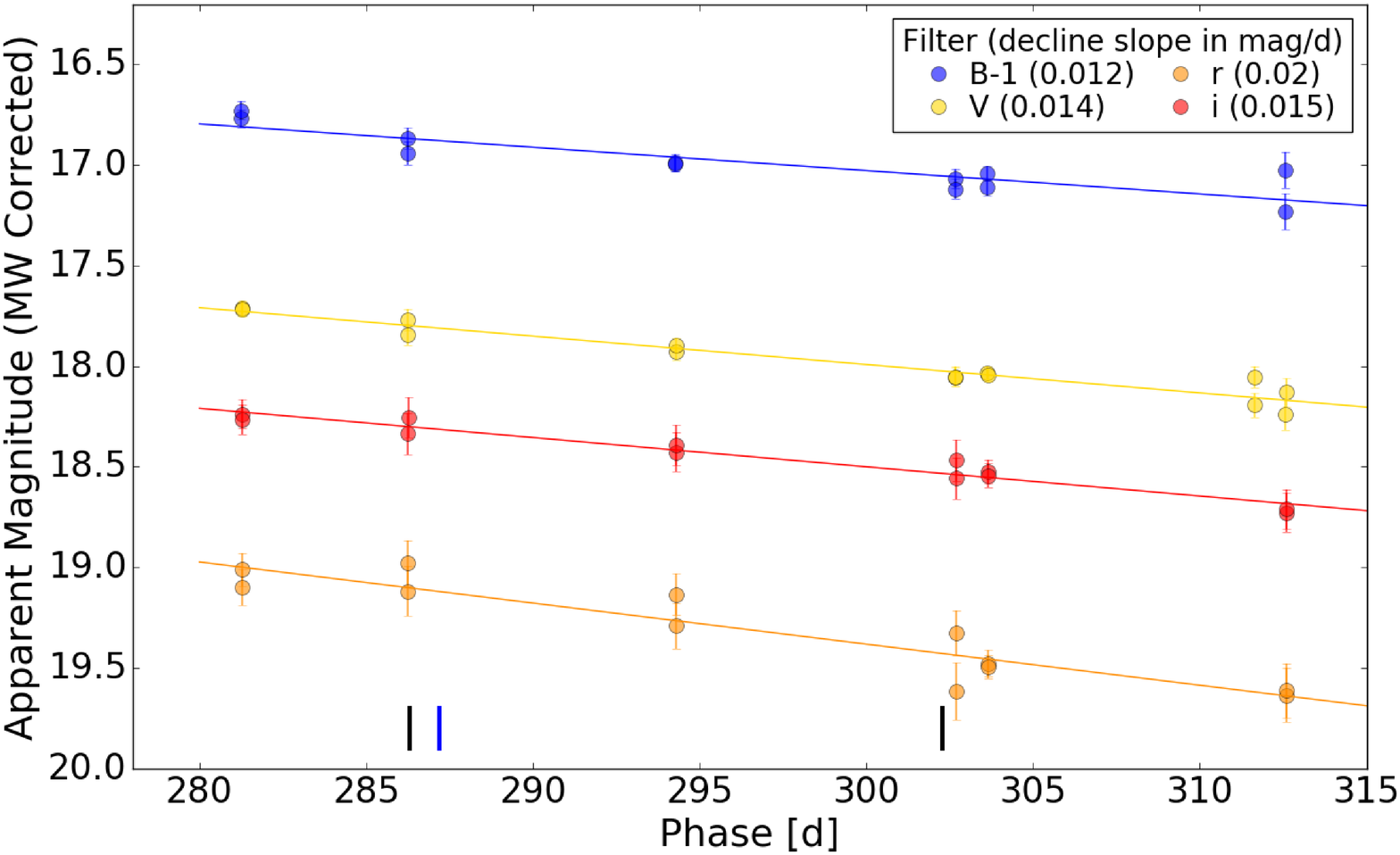}
\caption{{\it Top:} Multi-band SN~2017cbv light curve relative to $B$-band maximum light \cite[MJD 57841.07;][]{Hosseinzadeh17}.  Epochs with nebular spectroscopy in the optical (black) and NIR (blue) are marked in the lower right; the NIR and first optical spectrum  blend on this scale.  
Also plotted is the scaled SN~2011fe light curve at epochs greater than +80 d  \citep[][]{Graham15}.
 {\it Bottom:} A zoom in on the linear fits to the late-time light curve of SN~2017cbv, with decline slope given in the legend in $\rm mag\ d^{-1}$.}
\label{fig:lc}
\end{figure}

In order to flux calibrate our nebular spectra, we use the $BVri$ light curve obtained by Las Cumbres Observatory's network of 1-m telescopes \citep{Brown_2013}, which we will present in more detail in an upcoming publication (Sand et al., in prep.).  The images were reduced using \texttt{lcogtsnpipe}, a PyRAF-based photometric reduction pipeline \citep{Valenti_2016}, corrected for Milky Way (MW) extinction along the line of sight using $E(B-V)=0.1452$ mag \citep{Schlafly_2011}.  This light curve is presented in Figure~\ref{fig:lc}. The light curve up to +20 d after $B$-band maximum light was presented in \citet{Hosseinzadeh17}. For comparison, we scale the late-time light curve of normal Type Ia SN 2011fe \citep[taken from][]{Graham15} to match SN~2017cbv at phases $>+80$ d, represented by the semi-transparent lines in Figure~\ref{fig:lc}. In the lower panel of Figure \ref{fig:lc} we show a zoom in at late times, and our linear fits to the late-time decline rate in units of ${\rm mag\ d^{-1}}$. We find that SN~2017cbv is declining at a similar rate as SN~2011fe, with a slightly faster decline in the redder optical bands. We will 
present a more thorough photometric analysis in future work.

 \subsubsection{Distance to SN~2017cbv }

SN~2017cbv occurred in the outskirts of the nearby galaxy NGC~5643, which has a Tully-Fisher based distance measurement of $\mu$=31.14$\pm$0.40 mag, or $D$=16.9$\pm$3.1 Mpc \citep{Tully88}.  This distance is quite uncertain, but if true would imply that SN~2017cbv had an absolute magnitude of $M_{B}$=$-$20.0 mag for $D$=16.9 Mpc, and was significantly overluminous.  

To obtain an improved distance estimate, we used the light curve of SN~2017cbv directly along with the MLCS2k2 light curve fitter \citep{Jha07}.  We adopt a color excess associated with MW extinction of $E(B-V)=0.1452$ mag \citep{Schlafly_2011} as discussed above.    The resulting MLCS2k2 light curve fit is excellent with a host extinction of $A_V$=0.168$\pm$0.052, $\Delta$=$-$0.259$\pm$0.017 \citep[this parameter is the light curve stretch parameter in the MLCS2k2 methodology;][]{Jha07}, and $\mu$=30.45$\pm$0.09 mag (for $H_0$=72 km s$^{-1}$ Mpc$^{-1}$).  Although MLCS2k2 infers some host galaxy extinction, it is small and within the normal color dispersion of SN Ia.  Other observations infer negligible host extinction based on the narrow line equivalent widths of SN~2017cbv \citep{Ferretti17}, its color curve (Sand et al. in preparation) and its remote position in the host galaxy -- for this reason, we will assume no host exctinction throughout the rest of this work.

Given the MLCS2k2 distance modulus of $\mu$=30.45$\pm$0.09 mag ($D$=12.3$\pm$0.5 Mpc), and MW host extinction, this implies an absolute magnitude of $M_B$$\approx$$-$19.25 mag for SN~2017cbv, which is within expectations for a SN with its decline rate \citep[e.g.][]{Parrent14}.  We will use the revised distance modulus of $\mu$=30.45$\pm$0.09 mag ($D$=12.3$\pm$0.5 Mpc) throughout this work, but will also present results for the Tully-Fisher distance ($\mu$=31.14$\pm$0.40 mag) for completeness.

\subsection{Spectroscopy}\label{ssec:obs_spec}

\begin{figure*}
\begin{center}
\includegraphics[width=15cm]{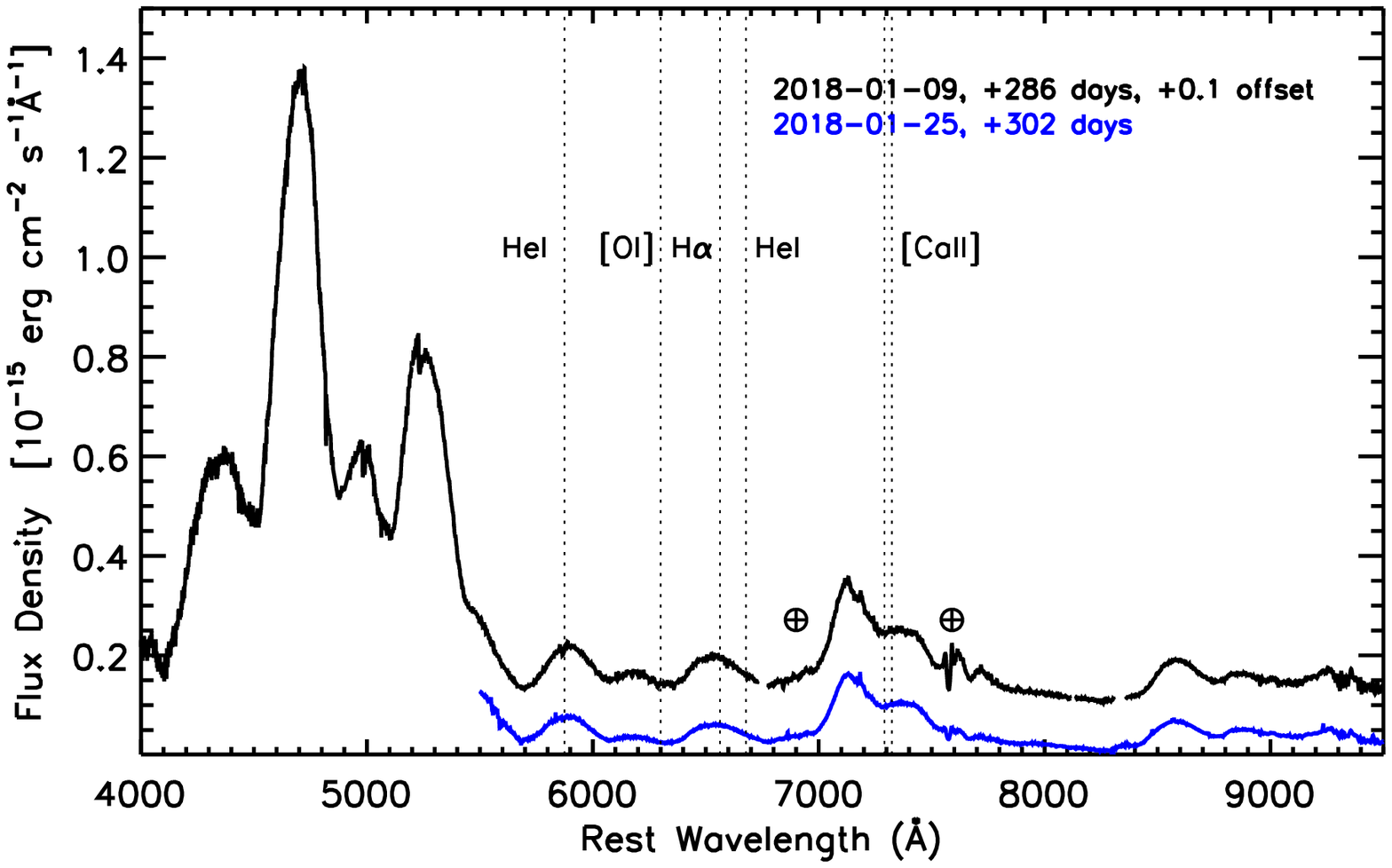}
\includegraphics[width=15cm]{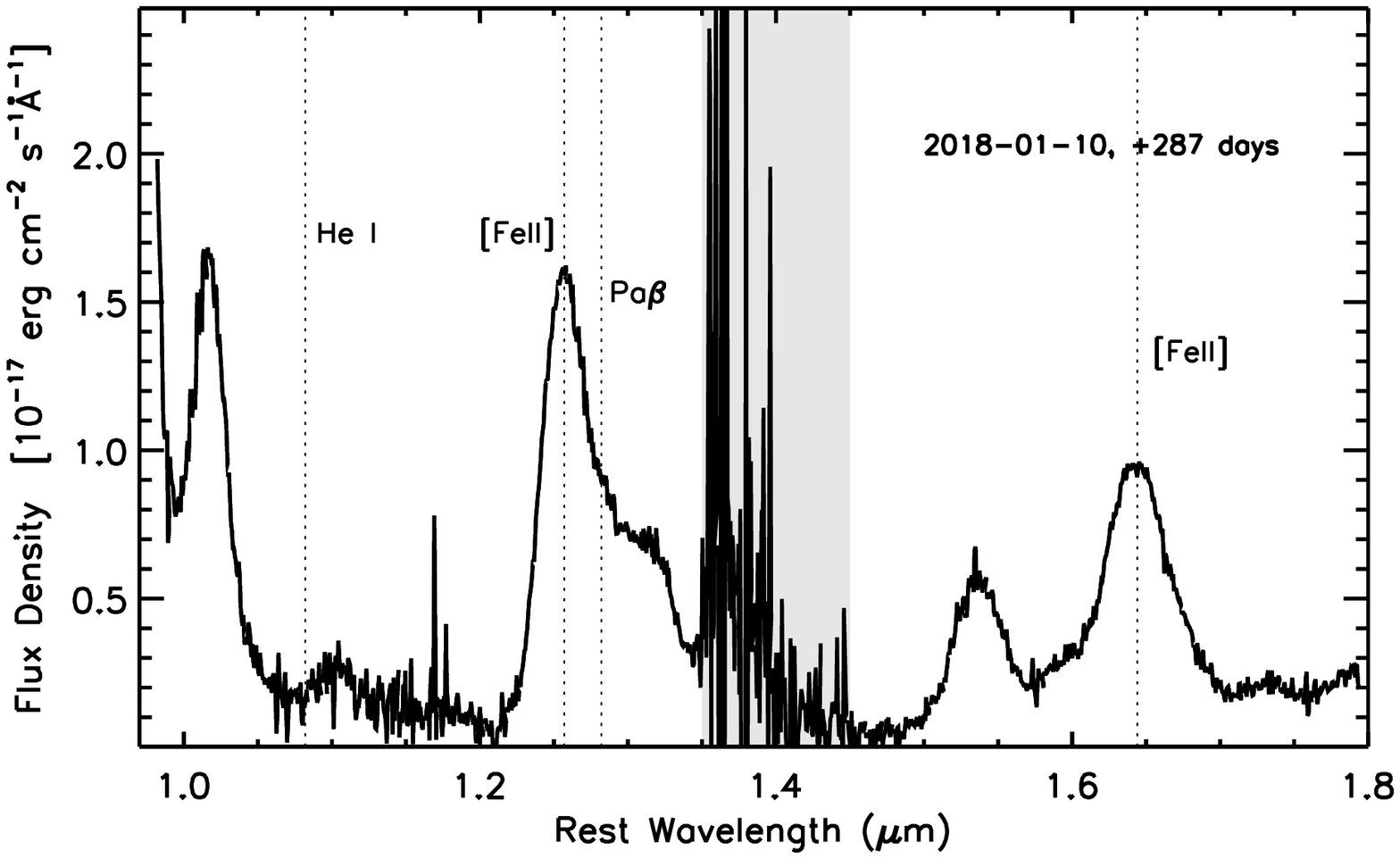}
\caption{Nebular-phase spectra of SN~2017cbv in the optical (top) and NIR (bottom), flux calibrated and corrected for MW extinction.  Phases refer to the number of days since $B$-band maximum. We label those wavelength regions where we search for narrow emission line features plausibly associated with companion star stripping.  In the optical spectrum, we mark the locations of the A- and B- band telluric features.  The gray vertical band in the NIR spectrum marks a region with strong telluric contamination. \label{fig:spec}}
\end{center}
\end{figure*}

We obtained two optical nebular spectra of SN\,Ia 2017cbv with the Gemini Multi-Object Spectrograph (GMOS; {\citealt{2004PASP..116..425H}}) on the Gemini South telescope on Cerro Pachon, Chile, as listed in Table \ref{tab:gmos} and shown in Figure~\ref{fig:spec}. 
An initial spectrum was taken on 2018 January 09 (UT), using the R400 and B600 grating in succession to get wavelength coverage from $\sim$4000--9500~\ \AA.  However, the sky conditions were variable, and one of the R400 exposures failed a quality assessment, leaving small gaps in the wavelength coverage on the red side.  A second spectrum was obtained on 2018 January 25 (UT), with a much longer exposure time of 4$\times$1200 $\rm s$ and utilizing only the R400 grating, with the goal of obtaining a very high signal-to-noise ratio spectrum in the region surrounding H$\alpha$.   We will focus on this second spectrum in Section~\ref{sec:ana} when we place limits on the mass of H-rich stripped material associated with SN~2017cbv.


The optical spectra were processed with the Gemini {\sc IRAF} package: the raw two-dimensional (2D) spectra were flat-fielded, bias-subtracted, and trimmed; cosmic-ray rejection was performed; wavelength calibration was determined using the Cu-Ar  arc lamp; and sky-subtracted one-dimensional (1D) apertures were extracted from the processed frames. The sensitivity function was determined from standard-star observations, and applied to flux calibrate our data. We also used the standard-star spectra to remove telluric features, although some residuals remain. 

We correct the spectra for MW extinction along the line of sight
using the extinction model of \cite{1999PASP..111...63F}. The host galaxy extinction is negligible based on the narrow line equivalent widths of SN~2017cbv \citep{Ferretti17}, along with its color curve (Sand et al. in preparation).  To place the optical spectra on an absolute scale, we rescale each spectrum to match the MW extinction-corrected $r$-band magnitude at  the nearest epoch -- this was $r$=19.05 mag for the 2018 January 9 (UT) epoch, and $r=19.47$ mag for the 2018 January 25 (UT) epoch.   We can see from the bottom panel of Figure \ref{fig:lc} that there is some scatter in the late-time $r$-band photometry, although the trend is clear and individual points are not affecting the overall fit. If we instead use our linear fit to estimate the $r$-band magnitudes on the dates of our spectroscopic observations we find $r$=19.10 and $r$=19.43 mag, respectively. This indicates that our overall uncertainty in the flux calibration of our spectra is at the $\sim0.05$ mag level, this is surely subdominant in comparison to the  current uncertainties in the models predicting narrow emission line flux (see \S~\ref{sec:conclude}). Finally, we correct for the redshift of the host, $z=0.00399$.  These final spectra are shown in Figure~\ref{fig:spec}.

A spectrum of SN~2017cbv was also obtained with the Flamingos-2  near-infrared (NIR) spectrograph \citep{f2} on the Gemini South Telescope, on 2018 Jan 10 (UT).  The NIR data were acquired and reduced using methods presented in earlier studies \citep[e.g.,][]{Sand16}, using a standard ABBA technique, with the slit position near the parallactic angle.  An A0V star was observed near in time and position to the science data in order to make a telluric absorption correction, and to flux calibrate the spectra, following the methodology of \citet{Vacca03}.  We again corrected this spectrum for MW extinction, and an absolute flux calibration was obtained directly by matching the blue portion of the NIR spectrum with the optical GMOS spectrum taken the day prior.

All of the spectra presented in this work are available on WISeREP\footnote{http://wiserep.weizmann.ac.il} \citep{Yaron12}.

\section{The Search for Stripped Material}\label{sec:ana}

If the early blue bump in the light curve of SN~2017cbv was caused by the SN ejecta impacting a nondegenerate companion star, then models predict that the stripped hydrogen (or helium) rich material would be swept up and cause narrow emission lines of FWHM $\approx1000$ $\rm km\ s^{-1}$ at late times. SN\,2017cbv is well separated from its host galaxy and no features of host-galaxy hydrogen emission are seen in the optical spectra  \citep[see also the spectra presented in][]{Hosseinzadeh17,Ferretti17}. 
It is clear from Figure \ref{fig:spec} that no strong narrow emission is obvious in the late-time spectra (the broad underlying emission line at $\sim$6500\ \AA~is [\ion{Co}{3}]). We focus our analysis on a search for trace amounts of H$\alpha$ in the deep optical spectrum obtained on 25 January, 2018 (at +302 d).  
We also search for evidence of other emission lines such as \ion{He}{1}, [\ion{O}{1}], [\ion{Ca}{2}], [\ion{Fe} {2}] and Pa$\beta$; we mark the position of these lines in Figure~\ref{fig:spec}, and remind the reader that we are searching for narrow features, in contrast to the standard broad features seen in SN spectra.

\subsection{H$\alpha$ Search}\label{sec:halpha}

Our statistical constraints on the presence of a narrow H$\alpha$ emission feature uses the higher signal-to-noise ratio GMOS spectrum obtained on 25 January 2018 UT (+302 days).  Using past observations and modeling work as a guide \citep[e.g.,][]{Mattila05,Leonard07,Shappee13,Lundqvist15,Boty18}, we assume ${\rm FWHM}=1000$ $\rm km\ s^{-1}$ for the line width of H$\alpha$ emission, and a potential offset from the rest wavelength of up to 1000 $\rm km\ s^{-1}$ as well. 

Our methodology borrows from previous work \citep[in particular;][]{Leonard07,Shappee13}, but differs in several respects.  First we take the flux-calibrated, extinction and redshift-corrected spectra  and bin to the spectral resolution of the data, $\sigma \approx 3.5$ \AA. This resolution is a combination of the spectrograph configuration and the seeing, as the seeing was significantly smaller than the slit width. We used the $1.5\arcsec$ slit but the seeing was ${\rm FWHM} \approx 0.88\arcsec$ (from the Gemini DIMM report and the acquisition images). The GMOS platescale is $0.16\arcsec$ $\rm pixel^{-1}$, and using the R400 grating with a CCD binning of 2$\times$2 gives a dispersion of 1.5 \AA\ $\rm pixel^{-1}$. Combined, the spectral resolution of our data is $\sigma = 1.5\ {\rm \AA\ pixel^{-1}} \times 0.88\arcsec / 0.16\arcsec\ {\rm pixel^{-1}} / 2.35 = 3.5$ \AA.

\begin{figure}
\begin{center}
\includegraphics[width=9cm]{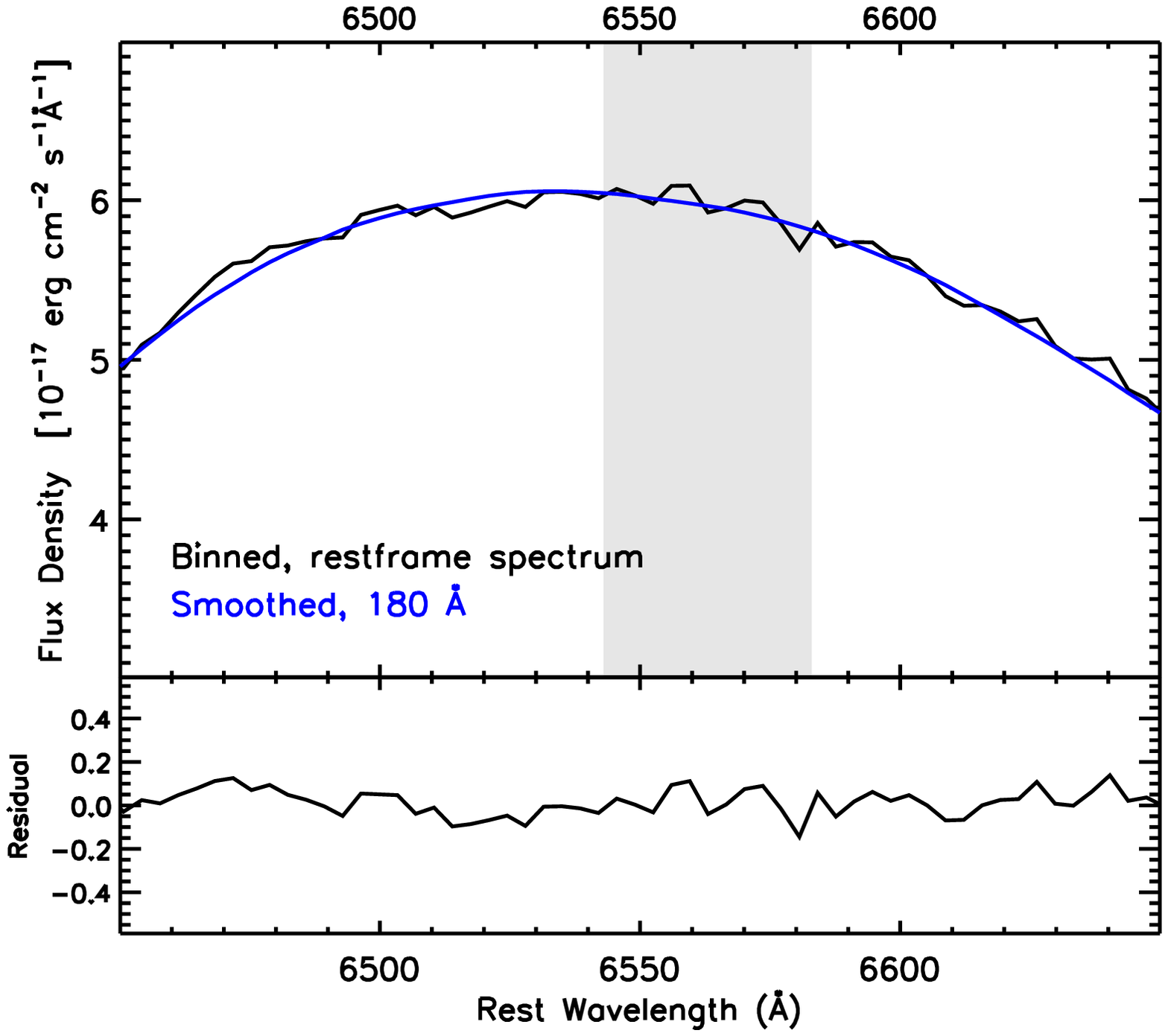}
\includegraphics[width=9cm]{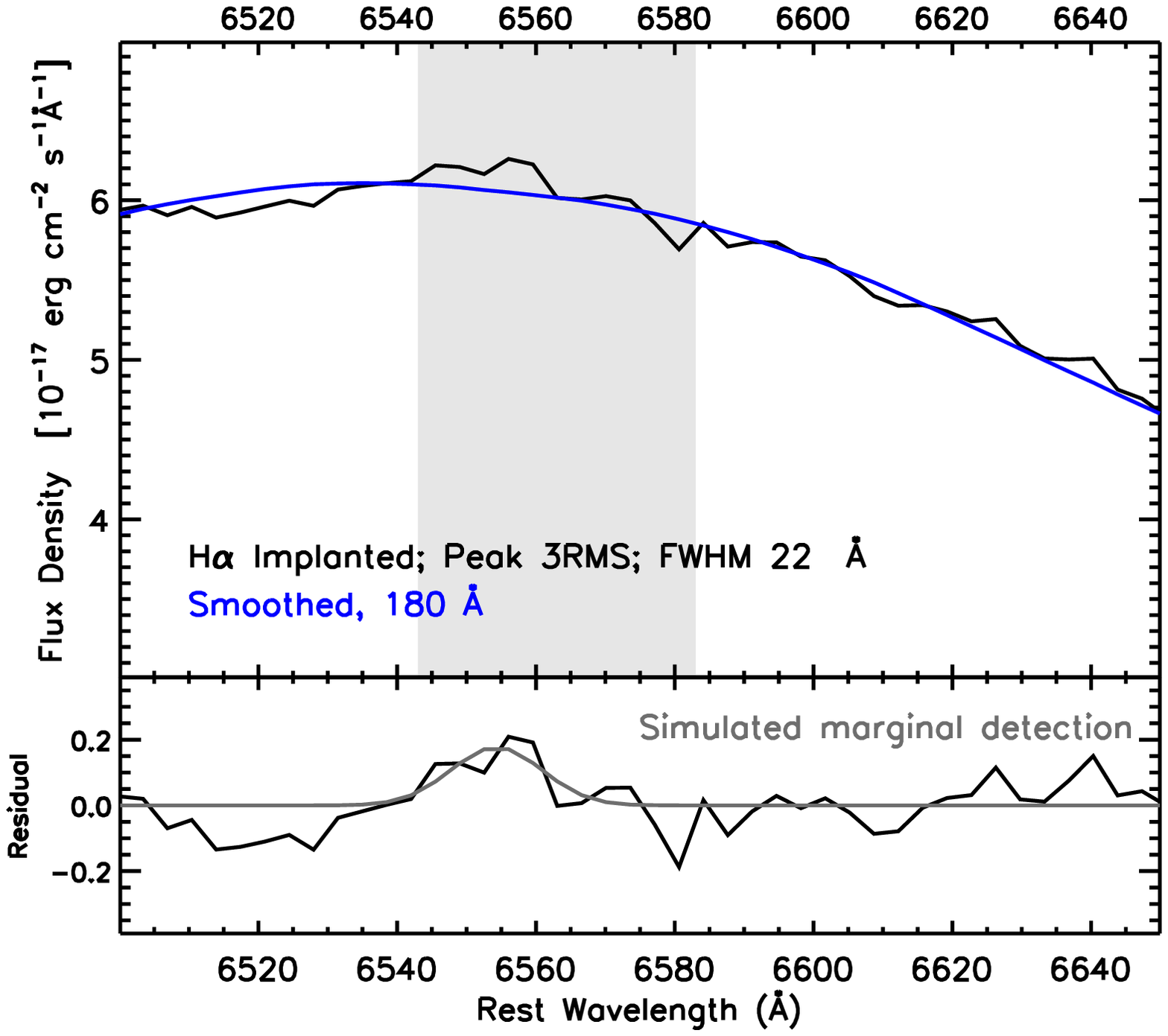}
\caption{Top -- The +302 d optical spectrum of SN~2017cbv in the region of H$\alpha$, illustrating our search technique for narrow emission line features.  The black line is the data binned to the resolution of the spectrum, while the blue shows the data smoothed by a second order Savitsky-Golay filter with a width of 180 \AA.  The bottom panel of the top figure shows the difference between the original spectrum and the smoothed version, with no discernible emission line-like residuals.  Bottom -- Similar to the top panel, but here we have implanted an H$\alpha$ feature at our estimated detection limit, with a peak value 3 times the RMS in the residual. A faint emission line feature is marginally recovered, with a flux within $\sim$30\% of that implanted feature.  The gray vertical band marks the region within 1000 km s$^{-1}$ of restframe H$\alpha$ in both plots.   \label{fig:halpha}}
\end{center}
\end{figure}

We then determine the continuum of the underlying broad [\ion{Co}{3}] emission by smoothing the spectrum on scales significantly larger than the expected narrower H$\alpha$ emission line signature, using a second-order Savitsky-Golay filter with a width of 180~\AA\ ($\sim 8200$ $\rm km\ s^{-1}$). This smoothing scale is slightly larger than that adopted by other work, but we found that smaller widths degraded our ability to detect and measure faint H$\alpha$ features implanted into our data, which we discuss below. The binned restframe spectrum and the smoothed spectrum in the region of H$\alpha$ are shown as black and blue lines respectively in the top panel of the top plot of Figure \ref{fig:halpha}. In the bottom panel of the top plot, we show the difference of the binned and smoothed spectra. Any features of the same scale as the expected H$\alpha$ signature ($\sim$1000 km s$^{-1}$) would manifest in this difference spectrum, but no such feature is apparent.

To estimate the maximum H$\alpha$ emission line that could remain undetected in our GMOS spectrum, we use a Gaussian profile with a width of ${\rm FWHM} \approx 1000$ $\rm km\ s^{-1}$ and a peak flux equal to three times the noise in our spectrum, where the noise is the root-mean-square (RMS) of our residual spectrum (e.g., as in \citealt{Graham17}). This results in a flux limit of 4.4$\times$10$^{-17}$ erg s$^{-1}$ cm$^{-2}$, and a luminosity limit of 8.0$\times$10$^{35}$ (1.5$\times$10$^{36}$) ergs s$^{-1}$ at a distance of D=12.3 (16.9) Mpc (Table~\ref{tab:lines}).

To test that a line with this estimated maximum H$\alpha$ emission is truly detectable, we implanted a Gaussian emission line (FWHM $\approx 1000$ $\rm km\ s^{-1}$) with an integrated flux equal to our limit (4.4$\times$10$^{-17}$ erg s$^{-1}$ cm$^{-2}$) into our binned restframe spectrum at a velocity of $-$500 km s$^{-1}$ \citep[a value typically seen in simulations;][]{Boty18}.  We then re-ran our analysis by smoothing with the 180~\AA\ Savitsky-Golay filter and differencing with the binned spectrum. This is illustrated in the bottom plot of Figure~\ref{fig:halpha}, where the implanted emission line is just discernible as an excess, given the spectrum's noise, at the left edge of the grey box. A Gaussian fit to the data recovers the peak flux to within $\approx$5\% and the total integrated flux to within $\approx$30\%. Although this is not a perfect recovery of the implanted feature, perfection would not be expected for an emission line at the detection limit. If such a feature was detected in a given data set, more detailed modeling would be needed in order to recover the maximum amount of information from the feature. Here we are simply demonstrating that a H$\alpha$ feature with a peak flux of 3 times the local RMS would be detectable in our data, and that fainter emission lines would be undetectable.

Previous works have used different statistical methods to derive upper limits on emission line flux, such as a formalism for calculating the 3$\sigma$ upper bound on the equivalent width \citep[see details in][]{Leonard01,Leonard07}. Flux limits using these statistical methods are a factor of $\sim$4 more stringent than those described here. As a test, we implanted simulated H$\alpha$ lines that represent the formal statistical flux limit based on equivalent width, but we were unable to recover them. We therefore adopt our verified statistical limit of a peak flux density equal to 3 times the local RMS.


\subsection{Search for Other Emission Lines}


Although nebular H$\alpha$ emission is the most discussed in the context of the single degenerate scenario, other hydrogen lines may be prominent \citep[e.g., Pa$\beta$;][]{Maeda14,Boty18}. If the companion is a helium star then \ion{He}{1} (at 5875, 6678 \AA~and 1.082 $\mu$m), [\ion{O}{1}]$\lambda$6300 and \ion{Ca}{2}$\lambda$$\lambda$7291,7324 are also plausible candidate lines for detecting the presence of a nondegenerate companion \citep{Liu13,Lundqvist13,Pan13,Lundqvist15,Boty18}.  Narrow emission lines associated with [\ion{Fe}{2}] at 1.257 and 1.644 $\mu$m \citep[not to be confused with the standard, very broad 1.644 $\mu$m feature seen in all SN~Ia NIR nebular spectra;][]{Diamond15,Diamond18} are also seen in some simulations \citep{Boty18}.  We searched our optical and NIR spectra for these narrow lines using the same statistical methodology described in \S~\ref{sec:halpha}, but do not detect any. 
For two NIR lines -- Pa$\beta$ and \ion{He}{1} $\lambda$1.082 $\mu$m -- we used a flux limit with a peak 4 times that of the RMS of the local spectral region (rather than three), because we could not confidently recover features of less significance.  We also used a peak flux limit with 4 times that of the RMS for the \ion{He}{1} $\lambda$5875 \AA~line, as this wavelength nearly coincides with a bright sky line at 5889 \AA~in the rest wavelength spectrum, and there is a noticeable residual in our spectrum.  We present our flux and luminosity limits in Table~\ref{tab:lines}.

Previous work in both SN~2011fe and SN~2014J has noted the presence of an emission feature at $\sim$7210 \AA, with a width approximately that expected from stripped gas \citep{Lundqvist15}, and we see the same feature in SN~2017cbv.  As this feature appears to be at approximately the same observed wavelength for all three SNe, we conclude (as did \citealt{Lundqvist15}) that this feature is an artifact, likely associated with a telluric line.

\section{Stripped Mass Limits} \label{sec:masslimits}

We have measured luminosity limits for the presence of narrow, low-velocity emission lines which are expected if a nondegenerate companion star was present in the progenitor system of SN~2017cbv.  No such emission was detected, and our limits are summarized in Table~\ref{tab:lines}.
Here we convert the line luminosity limits to limits on the mass of hydrogen, focusing on the H$\alpha$ line.  We also discuss helium mass constraints.

Most previous searches have used the results of 1D radiative transport models to relate nebular emission to the amount of stripped hydrogen mass \citep{Mattila05,Lundqvist13}.  Here we use the recent 3D radiation transport results of \citet{Boty18}, who present simulated SN\,Ia spectra at $200$ days after peak brightness that incorporate stripped material from a companion star (of solar abundance) mixed into the core of the ejecta.  These models were derived from the SN Ia ejecta-companion interaction simulations of \citet{Boehner17}. \cite{Boty18} find that narrow hydrogen (and helium) nebular emission should be even stronger than previously suggested, differing with the 1D results by roughly an order of magnitude.  
Models included main sequence, subgiant and red giant companion stars, with stripped masses between $\sim$0.2--0.4 $M_{\odot}$, and luminosities between $L_{H\alpha}$$\approx$4.5--15.7$\times$10$^{39}$ ergs s$^{-1}$ \citep[see Table~1 in ][]{Boty18}.  The H$\alpha$ line center is seen to vary with viewing angle ($\sim$10 \AA), but it is within the 1000 km s$^{-1}$ region examined in the previous section.

A comparison of our luminosity limits with the expected results from \citet{Boty18} rule out any of their companion interaction models, as our conservative $L_{H\alpha}$$<$8.0$\times$10$^{35}$ ergs s$^{-1}$ is over three orders of magnitude less than expectations.  More specifically, taking the best-fit companion interaction model of SN~2017cbv based on the early blue bump in the light curve, \citet{Hosseinzadeh17} inferred the companion to be a subgiant star at a binary separation of $\sim$60 $R_{\odot}$ from the white dwarf (although this is highly model dependent and should be treated with caution).  This would place SN~2017cbv's companion between the subgiant and red giant interaction models of \citet{Boty18}, corresponding to $L_{H\alpha}$$\approx$5$\times$10$^{39}$ ergs s$^{-1}$, which is clearly ruled out by our nebular spectra.

For a quantitative comparison to the models' H$\alpha$ emission line luminosities, we must first apply a correction factor to account for the fact that the models are derived for $+200$ d after peak brightness, but our deep GMOS spectrum was obtained at $+302$ d after peak. \cite{Boty18} find that the H$\alpha$ luminosity is a relatively constant fraction of both the \ion{Fe}{3} $\lambda4658$ nebular feature and the bolometric luminosity, both of which decline approximately linearly with time during the nebular phase of SNe\,Ia -- by about a factor of four between 200 and 300 d \citep{2015MNRAS.450.2631M}. To convert the models' expected H$\alpha$ luminosity at $+200$ d to $+302$ d we simply divide by 4, as do \cite{Boty18} in their re-analysis of the H$\alpha$ limit for SN\,2011fe based on a spectrum at $296$ days after peak brightness \citep{Shappee13}. After this correction, we would still expect an H$\alpha$ luminosity of $L_{H\alpha}$$\gtrsim$1$\times$10$^{39}$ ergs s$^{-1}$.

 If the mass of stripped hydrogen is lower, for instance if the companion separation is larger than that considered in the \citet{Boty18} models or if an asymmetric explosion caused a smaller portion of the ejecta to interact with the companion, then the expected $L_{H\alpha}$ would be concomitantly lower.  To approximate this effect, \citet{Boty18} varied the hydrogen density in their fiducial main sequence companion model, finding a quadratic fitting formula that related the stripped hydrogen mass and the H$\alpha$ luminosity (see their Equation 1, but note that the coefficients had been published backwards and the proper form is $\log_{10}(L_{\rm H\alpha}) = -0.2M_1^2 + 0.17M_1 +40.0$, where $M_1 = \log_{10}(M_{\rm st}/M_{\odot})$, and $M_{\rm st}$ is the stripped mass).  Applying this function to our own H$\alpha$ luminosity limit, after we have converted it from $+302$ days to $+200$ days after peak brightness ($\log_{10}(L_{\rm H\alpha}) = 36.33$), implies that the maximum amount of undetected stripped hydrogen has a mass of $\sim$1$\times$10$^{-4}$ M$_{\odot}$. However, we caution that this hydrogen mass limit is based on the extrapolation of the \citet{Boty18} simulations, and refer the reader to that work for important caveats.  

For comparison with past work, we also use the models of \citet{Mattila05} to make an estimate of the maximum hydrogen mass that might remain undetected. \citet{Mattila05} show that $0.5$ $\rm M_{\odot}$ of hydrogen would produce an emission line with peak intensity of $3.36\times 10^{35}$ $\rm erg\ s^{-1}\ \AA$ at $380$ days after peak brightness. By scaling this intensity to the epoch of our GMOS spectrum at $\sim 300$ days after peak (again using the factor of 4 decline in flux every 100 days described above), and to the distance of SN~2017cbv, this would correspond to an emission line with a peak flux density of $6.0\times 10^{-17}$ $\rm erg\ s^{-1}\ cm^{-2}\ \AA$ (on top of the underlying SN nebular emission), or an equivalent width of $W_{\lambda}(0.5~{\rm M_{\odot}})\approx23.1$ \AA. Our statistical maximum from Table \ref{tab:lines}, however, has an equivalent width of $W\approx0.73$ \AA. The models of \citet{Mattila05} show a linear scaling between the mass of hydrogen and the equivalent width of the emission line, and so our limit on the mass of hydrogen, $M_H$, is:

\begin{equation}
M_H \lesssim \frac{0.73~{\rm \AA}}{23.1~{\rm \AA}} \times 0.5~{\rm M_{\odot}} \approx 0.001~{\rm M_{\odot}}
\end{equation}

\noindent This is an order of magnitude less constraining than our limit based on the new models of \cite{Boty18}. For comparison, \citet{Shappee13} used this same method to derive a similar lower limit of $M_H \lesssim 0.001$ $\rm M_{\odot}$ for SN~2011fe.

\subsection{Helium mass limits}

In addition to the hydrogen-rich companion scenario discussed thus far, helium stars have long been considered viable white dwarf companions in the progenitor SN Ia system \citep[e.g][]{Iben84}, with recent hydrodynamic interaction models predicting $\approx$0.0024--0.06 $M_\odot$ of unbound helium \citep{Pan12,Liu13}.  Connecting helium line luminosities to stripped mass is less straightforward than for the case of hydrogen, as fewer radiative transport simulations have been done.  Nonetheless,  \citet{Boty18} attempted to model the effect by simply replacing hydrogen with helium in their simulation.  If we assume the relation between stripped helium mass and helium luminosity falls off with the same quadratic form as hydrogen, we infer a limit on the amount of undetected stripped helium to be $\sim$5$\times$10$^{-4}$ M$_{\odot}$ (using the He I $\lambda$10830 line).  Other works have suggested that helium would be difficult to detect in nebular phase optical spectra, and instead used the [\ion{O}{1}]$\lambda$6300 and \ion{Ca}{2}$\lambda$$\lambda$7291,7324 lines as potential helium tracers \citep{Lundqvist15}.  Using the methodology of \citet{Lundqvist15}, and their published helium mass limits for SN~2011fe and SN~2014J based on nebular spectra, we infer a scaled helium mass limit of $\lesssim$1$\times$10$^{-3}$ M$_{\odot}$ from our SN~2017cbv spectra.  In either scenario, our helium mass limits are more restrictive than current helium companion models predict \citep[e.g.,][]{Pan12,Liu13}, although we caution that more observational and modeling work needs to be done.  We  note in passing that helium stars are also candidate companions to SNe~Iax systems \citep[for a recent review see][]{Jha17}, with direct observational evidence in some circumstances \citep[e.g.,][]{Foley13,McCully14}, and future high signal-to-noise ratio late time spectra may be another route to constraining their progenitor systems.


\begin{table*}
\begin{center}
\begin{tabular}{lccc}
\hline
\hline
Emission & Flux Limit   & Luminosity Limit (D=12.3 Mpc) & Luminosity Limit (D=16.9 Mpc)   \\ 
Line   & (10$^{-17}$ erg s$^{-1}$ cm$^{-2}$) &   (10$^{36}$ ergs/s)         & (10$^{36}$ ergs/s)            \\
\hline
H$\alpha$ $\lambda$6563 & 4.4 & 0.8        & 1.5     \\ 
He I $\lambda$5875 & 18.6 & 3.4 & 6.4 \\
He I $\lambda$6678 & 8.3 & 1.5 & 2.8 \\
$[$OI$]$ $\lambda$6300 & 8.6 & 1.5 & 2.9 \\
$[$CaII$]$ $\lambda$$\lambda$7291,7324 & 26.0 & 4.7 & 8.9 \\
He I $\lambda$10830 & 5.0 & 0.9    & 1.7      \\
$[$FeII$]$ $\lambda$12570 & 3.9 & 0.7 & 1.3 \\
Pa$\beta$ $\lambda$12820 & 3.2 & 0.6 & 1.1 \\
$[$FeII$]$ $\lambda$16440 & 2.3 & 0.4 & 0.8 \\
\hline
\end{tabular} 
\caption{Emission line flux and luminosity limits.  All implanted lines have peak fluxes corresponding to three times the local RMS with a FWHM=1000 km s$^{-1}$ (see Section 3.1 for details), except for the Pa$\beta$, He I $\lambda$10830 and He I $\lambda$5875  lines, where we used a peak flux of four times the local RMS. We convert the flux limits to luminosity for the two estimates for host galaxy distance (see Section \ref{sec:intro}).}
\label{tab:lines}
\end{center}
\end{table*}

\section{Conclusions \& Summary}\label{sec:conclude}


In this work we have analyzed the late-time spectra of SN~2017cbv, a SN\,Ia that exhibited a distinctive ``blue bump" in its early-time light curve.  A plausible interpretation is that this bump was a signature of the SN's ejecta interacting with a nondegenerate binary companion star. We find none of the predicted late-time narrow line emission that should be generated by the stripped material from the nondegenerate companion, and place an upper limit on the mass of hydrogen ($M_{\rm H} \lesssim 0.0001$ $\rm M_{\odot}$) and helium ($M_{\rm He} \lesssim 0.0005$ $\rm M_{\odot}$) that might remain undetected in the progenitor system.   These results are contingent on the necessary 3D radiative transport modeling that connects our luminosity limits to inferred mass limits of hydrogen and helium, and ultimately our mass limits are tied to the implantation of lower density material directly into the \citet{Boty18} models.  Models with 1D radiative transport yield mass limits roughly an order of magnitude higher than that quoted above. The presented mass limits are strong, and at face value argue against a nondegenerate companion in Roche lobe overflow as the progenitor of SN~2017cbv. Nonetheless, given the caveats above and the need for further simulations with varying companion types, separations, explosion energies and envelope masses, we cannot draw definitive conclusions on the SN~2017cbv progenitor system.



There are alternative mechanisms that may have yielded SN~2017cbv's early light curve.    One option is that the early-time blue bump of SN~2017cbv is a phenomenon related directly to some property of the white dwarf star and/or its explosion. For example, \cite{Piro16} show that mixing of radioactive $^{56}$Ni into the outer layers of the exploding white dwarf might also be the root cause of an early blue bump, especially if there is some circumstellar material in the system. Similarly, the early SN~Ia light curve models of \citet{Noebauer17} suggest that an early blue bump may be visible in sub-Chandrasekhar double detonation explosions. For Chandrasekhar-mass models a blue bump might occur naturally at early times if the photosphere is in the carbon layer, which has a lower opacity in the UV \citep[][although we note that SN~2017cbv did not have a UV bump]{2017arXiv170703823G}.  One advantage of the original \cite{Kasen10} single degenerate companion interaction light curves is that they were truly predictive -- no such early light curve features had been observed at the time of those simulations, but they motivated the current generation of fast cadence SN searches, such as DLT40, which are now revealing light curves with remarkably similar morphologies to that predicted.  We will further explore and test the scenarios mentioned above in our future analysis of the full SN~2017cbv data set.



\acknowledgments

Research by DJS and SW is supported by NSF grants AST-1821987 and 1821967.
The work of CM, GH, and DAH
is supported by US National Science Foundation (NSF) grant AST–
1313484. E. Y. H. and S. K. acknowledge the support provided by the National Science Foundation under Grant No. AST-1613472 and by the Florida Space Grant Consortium.

This research makes use of observations from the Las Cumbres Observatory. 
Based on observations obtained at the Gemini Observatory under programs GS-2017B-Q-14 (PI: Howell) and GS-2017B-Q-58 (PI: Sand).  Gemini is operated by the Association of Universities for Research in Astronomy, Inc., under a cooperative agreement with the NSF on
behalf of the Gemini partnership: the NSF (United States), the National
Research Council (Canada), CONICYT (Chile), Ministerio de Ciencia, Tecnolog\'ia e Innovaci\'on Productiva (Argentina), and Minist\'erio da Ci\^encia, Tecnologia e Inova\c{c}\~ao (Brazil). The data were processed using the Gemini IRAF package. We thank the queue service observers and technical support staff at Gemini Observatory for their assistance.

This research has made use of the NASA/IPAC Extragalactic Database (NED) which is operated by the Jet Propulsion Laboratory, California Institute of Technology, under contract with NASA.

%

\vspace{5mm}
\facilities{Las Cumbres Observatory, Gemini South (GMOS, Flamingos-2)}


\software{
astropy \citep{2013A&A...558A..33A,astropy},  
          }

\end{document}